\documentclass{article}

\PassOptionsToPackage{numbers, sort, compress, comma, square}{natbib}
\usepackage[final]{neurips_2023}

\usepackage[utf8]{inputenc} 
\usepackage[T1]{fontenc}    
\usepackage{hyperref}       
\usepackage{url}            
\usepackage{booktabs}       
\usepackage{amsfonts}       
\usepackage{nicefrac}       
\usepackage{microtype}      
\usepackage{xcolor}         
\usepackage{graphicx}
\usepackage{subcaption}
\usepackage{tikz}
\usepackage{tikz-3dplot}
\usetikzlibrary{arrows.meta}
\usepackage{graphicx}
\usepackage{wrapfig}
\usepackage{amsmath}
\title{Unrolling Virtual Worlds for Immersive Experiences}

\author{
  Alexey Tikhonov\\
  Inworld.AI \\
  \texttt{altsoph@gmail.com} \\
  \And
  Anton Repushko \\
  Independent Researcher \\
  \texttt{anton@repushko.com} \\
}

\begin{document}
\maketitle

\raggedbottom

\begin{abstract}
  This research pioneers a method for generating immersive worlds, drawing inspiration from elements of vintage adventure games like Myst and employing modern text-to-image models. We explore the intricate conversion of 2D panoramas into 3D scenes using equirectangular projections, addressing the distortions in perception that occur as observers navigate within the encompassing sphere. Our approach employs a technique similar to "inpainting" to rectify distorted projections, enabling the smooth construction of locally coherent worlds. This provides extensive insight into the interrelation of technology, perception, and experiential reality within human-computer interaction.
\end{abstract}

\section{Motivation}

In the field of human-computer interaction, the concept of immersive technologies has a rich history, dating back to the 1830s with the creation of the first stereoscopes during the early days of photography. We will understand the term immerse as "Technology that blurs the line between the physical, virtual, and simulated worlds, thereby creating a sense of immersion"\citep{lee2013}.

These technologies have evolved to serve many purposes, acting as mediums for education, psychotherapy, physiotherapy, interactive simulations, and entertainment \citep{suh}. The phenomenon of immersion is also well-established in modern philosophy, which only increases researchers' interest in this topic \citep{merleau-ponty_phenomenology_1945, bostrom_simulation_2003}.
One of the key and most advanced areas of immersive technologies is virtual environment creation. The advent of virtual reality headsets has enabled users to attain remarkable levels of immersion in virtual worlds, even fostering a market area for startups focusing specifically on world creation \citep{blockadelabs}.

Inspired by vintage games like Myst \citep{myst}, where immersion into the world was achieved through interconnected scenes creating a coherent world, we propose a novel method for developing consistent environments. This approach combines contemporary text-to-image models with stereometric transformations to innovate environmental generation, presenting a sophisticated strategy for crafting immersive spaces.

\section{Approach}

Despite their primary function being text-to-image conversion, modern text-to-image models often feature additional modes, such as text+image -> image. We utilized the fine-tuned StableDiffusion v1.5 model \citep{latentlabs360}, capable of generating panoramas in equirectangular projection. This projection, a 2x1 rectangle, can be seamlessly converted into a spherical panorama by using open-source tools \citep{pannellum}.

\begin{figure}[htbp]
    \centering
    \begin{subfigure}{0.3\textwidth}
        \centering
        \includegraphics[width=\linewidth]{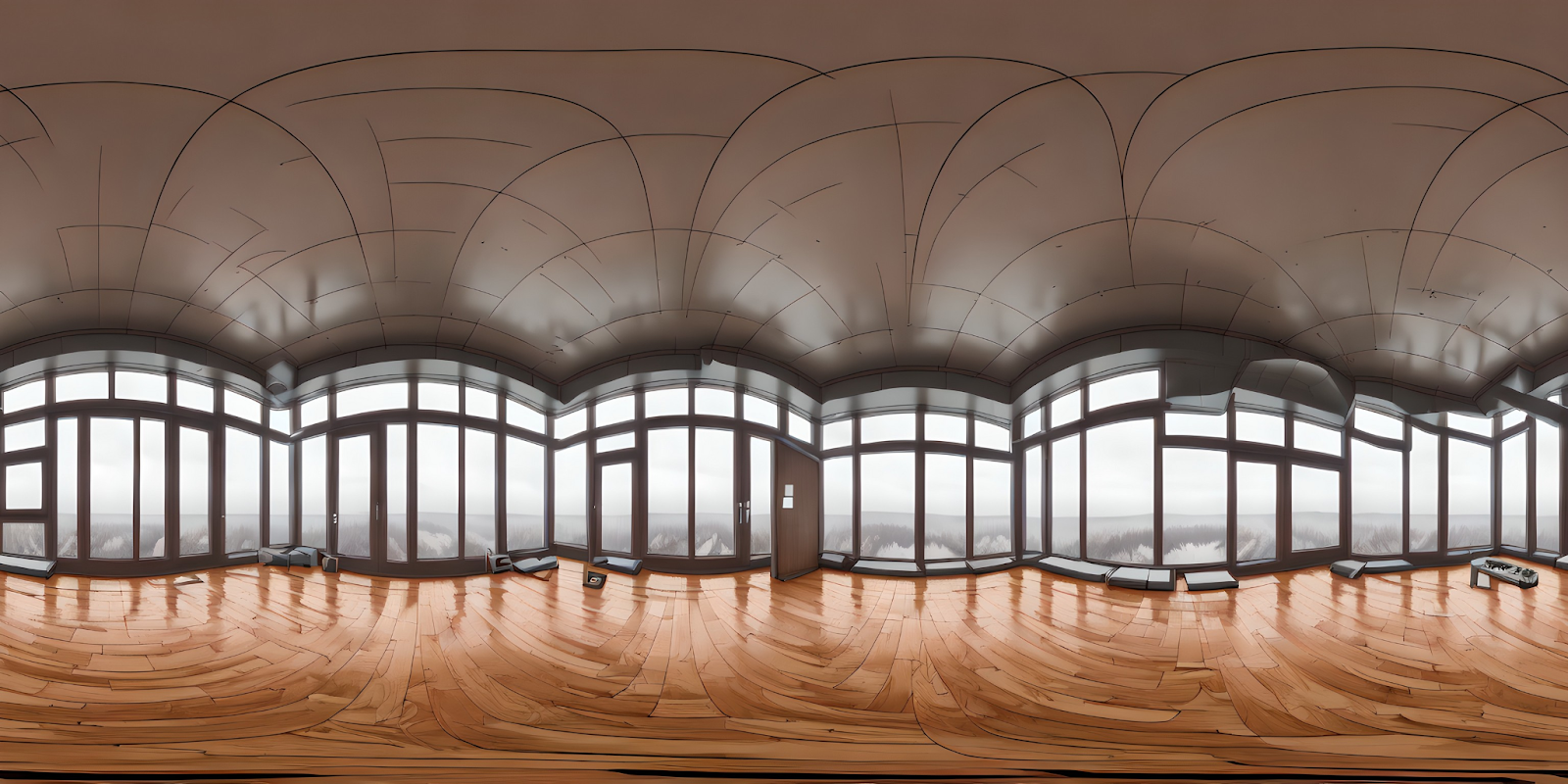}
        \caption{Initial panorama}
    \end{subfigure}
    \hfill
    \begin{subfigure}{0.3\textwidth}
        \centering
        \includegraphics[width=\linewidth]{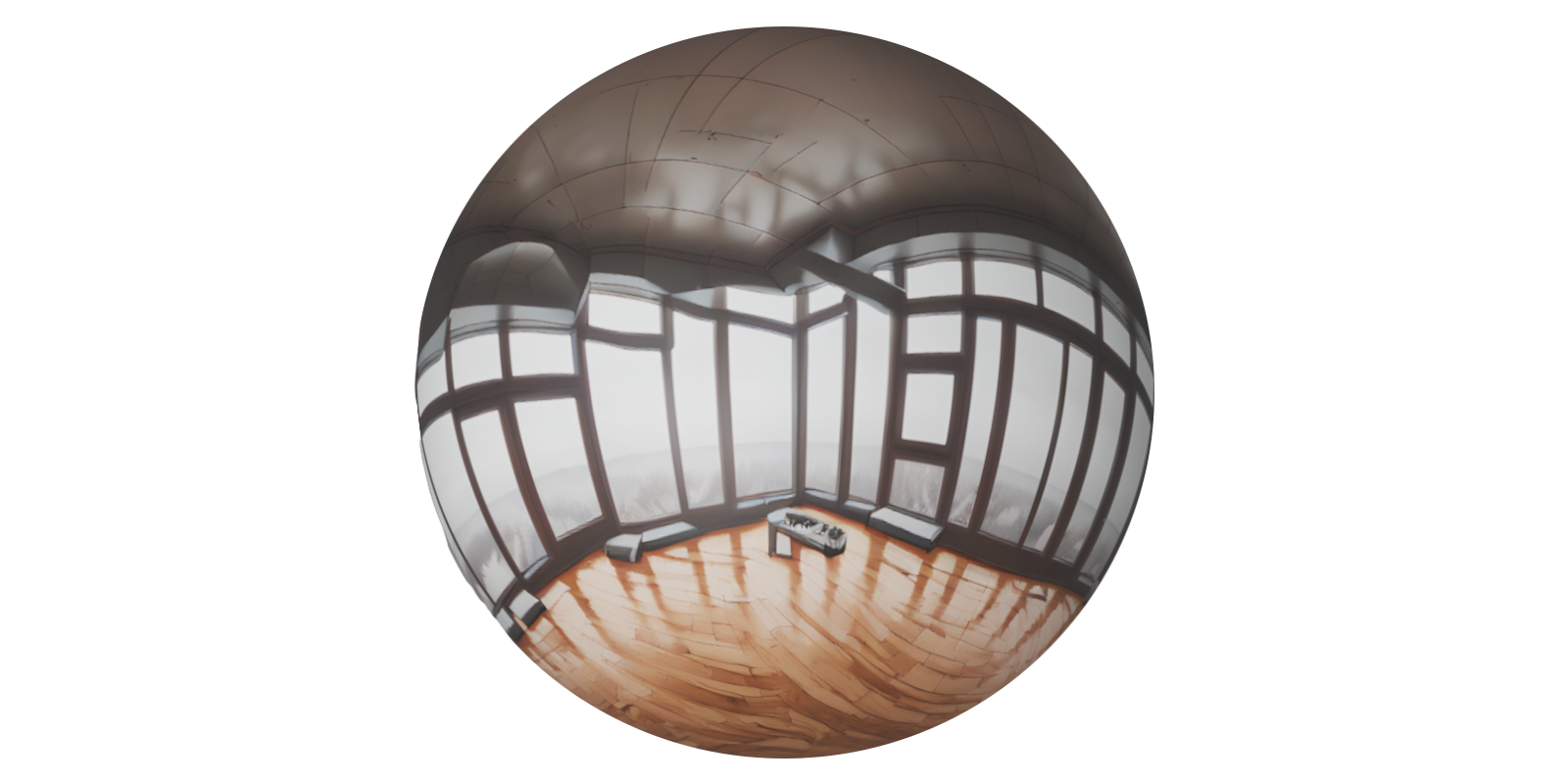}
        \caption{Panorama mapped on sphere}
    \end{subfigure}
    \hfill
    \begin{subfigure}{0.3\textwidth}
        \centering
        \includegraphics[width=\linewidth]{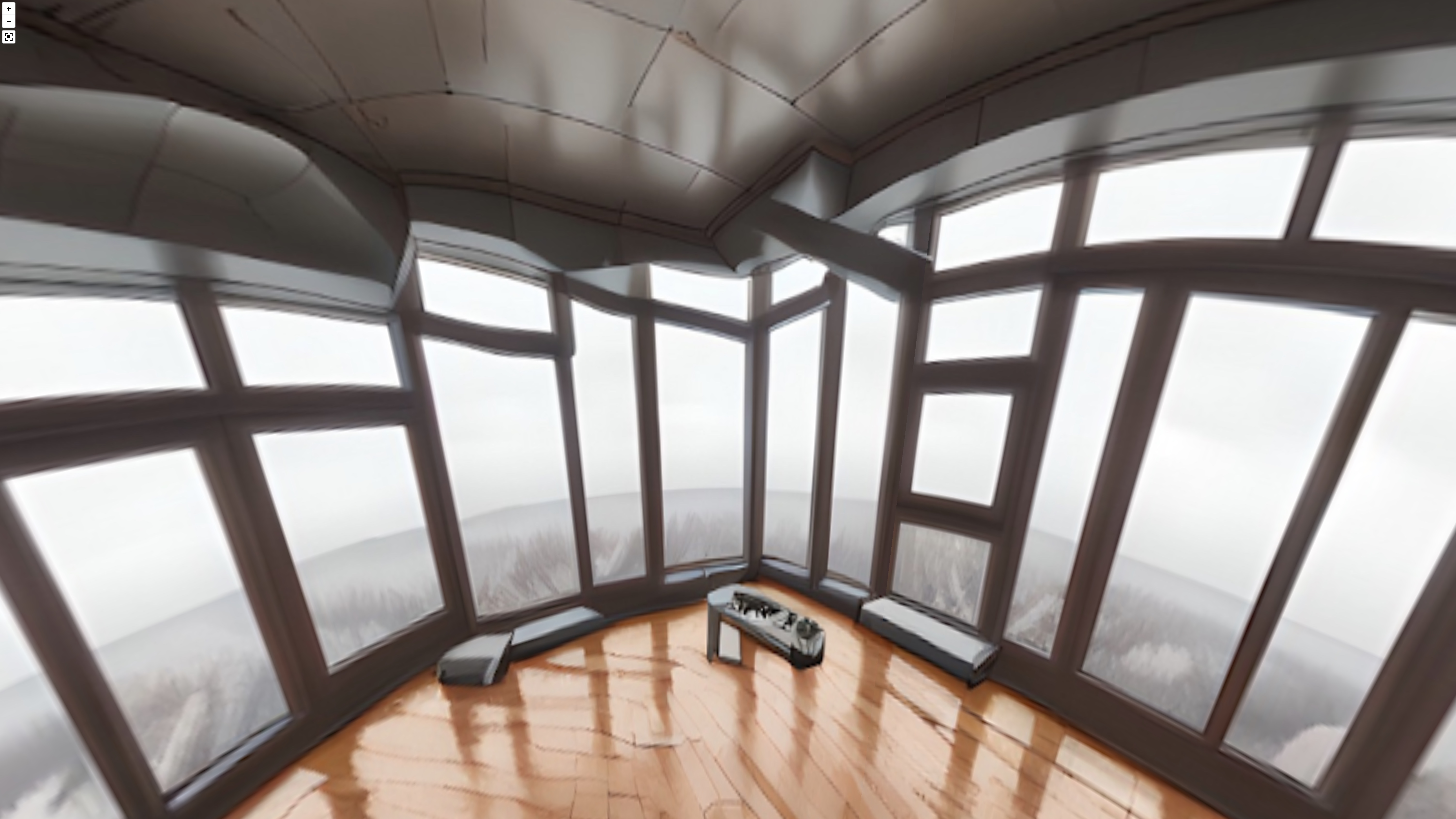}
        \caption{View inside the sphere}
    \end{subfigure}
    \caption{Generating scene from the panorama in equirectangular projection.}
    \label{fig:three_images}
\end{figure}
Thus, we can generate our initial scene using a prompt with an environment description. This spherical panorama already imparts a sense of being inside the world. The phenomenon of immersion within the panorama is well-known and heavily utilized by various entities, including game developers and museums, as a means to situate the viewer within a specific context and environment.

This immersive magic begins to unravel as we move in any direction away from the center. If we do so within this static spherical panorama, we will observe visual distortion, altering our perception of the panorama.

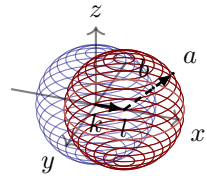
\begin{wrapfigure}{r}{0.4\textwidth}
  \centering
    \tdplotsetmaincoords{60}{110} 
    \begin{tikzpicture}[tdplot_main_coords, scale=0.4]
    
        \draw[thick,->, opacity=0.5] (-3,0,0) -- (3,0,0) node[anchor=north east, opacity=1]{$y$};
        \draw[thick,->, opacity=0.5] (0,-3,0) -- (0,3,0) node[anchor=north west, opacity=1]{$x$};
        \draw[thick,->, opacity=0.5] (0,0,0) -- (0,0,3) node[anchor=south, opacity=1]{$z$};
        
        \foreach \a in {0,10,...,350} {
            \pgfmathsetmacro{\R}{2*cos(\a)} 
            \pgfmathsetmacro{\z}{2*sin(\a)} 
            \draw[canvas is xy plane at z=\z,blue!50!black, opacity=0.3] (0,0) circle (\R);
        }
        \draw[thick, color=blue!25, opacity=0.15] (0,0,0) -- (2,0,0) arc[start angle=0, end angle=360, radius=2] -- cycle;

    
        \foreach \a in {0,10,...,350} {
            \pgfmathsetmacro{\R}{2*cos(\a)} 
            \pgfmathsetmacro{\z}{2*sin(\a)} 
            \draw[canvas is xy plane at z=\z,red!50!black, opacity=0.8] (0,1) circle (\R);
        }
        \draw[thick, color=red!25, opacity=0.4] (0,1,0) -- (2,1,0) arc[start angle=0, end angle=360, radius=2] -- cycle;
        
        \pgfmathsetmacro{\Ax}{2*cos(60)} 
        \pgfmathsetmacro{\Ay}{2*sin(60)} 
        \pgfmathsetmacro{\Az}{2*sin(60)}
        \fill (0,\Ay+1,\Az) circle (1.5pt); 
        \draw[densely dashed,-{Latex[length=2mm]}, line width=1pt] (0,1,0) -- (0,\Ay+1,\Az) node[anchor=south west]{$a$};
    
        \fill (0,\Ay+0.3,\Az*0.6) circle (1.5pt); 
        \node[anchor=south east] at (0, \Ay+0.5, \Az*0.65) {$b$};
    
        \fill (0,0,0) circle (1.5pt); 
        \node[anchor=north] at (0,0,0) {$k$}; 
    
        \fill (0,1,0) circle (1.5pt); 
        \node[anchor=north] at (0,1,0) {$l$}; 
    
        \draw[-{Latex[length=2mm]}, color=black, opacity=1, line width=1pt] (0,0,0) -- (0,1,0);

    \end{tikzpicture}
\caption{Movement within the scene generates a new expected panorama to maintain the feeling of immersion.}
\end{wrapfigure}

We have derived a formula to obtain this distorted image following the viewer's movement (can be found in the Appendix). In Figure 2, the blue sphere represents the current scene, and the red sphere represents the new expected panorama after movement from the old center \(k\) to the new center \(l\).

For every point ``$a$'' on the surface of the new sphere, we can find an intersection point ``$b$'' on the old sphere, determined by the radius from the new sphere center to point ``$a$''. Since point ``$a$'' represents a pixel in a new 2D panorama, we will use the pixel representing point ``$b$'' in the previous panorama to determine the color of this pixel. This is how we obtain the distorted panorama image.

After obtaining the distorted projection, we can remove this distortion using the model from the initial scene generation step, along with the initial prompt. Modern neural networks tend to "reconstruct" inputs, thereby enhancing their likelihood. We attribute this phenomenon to the networks' ability to learn the manifold of plausible objects, grounding various noisy or distorted objects onto this manifold and projecting them to the nearest suitable region. This can be interpreted as a form of spontaneous denoising.

In this methodology, the distorted projection is passed through the network utilizing the same text prompt but coupled with the reconstructed input image from the preceding step. This technique is essentially similar to "inpainting"; however, its application in this unique form for amending distorted projections is a novel exploration in our study. Pairs of the distorted and restored images can be viewed in Figures 4 and 5 in the Appendix.

Transitions between neighboring scenes occur smoothly and seamlessly, despite a noticeable pattern of accumulating errors and hallucinations over iterations. To immerse in a fully realized world, drawing inspiration from old adventure games, a grid of scene-image-panorama generations can be created. Examples of such worlds can be found in the demo \footnote{\url{https://altsoph.github.io/immersive_spaces}} and in Figure 6 in the Appendix.

\begin{figure}[htbp]
  \centering
  \includegraphics[width=1\textwidth]{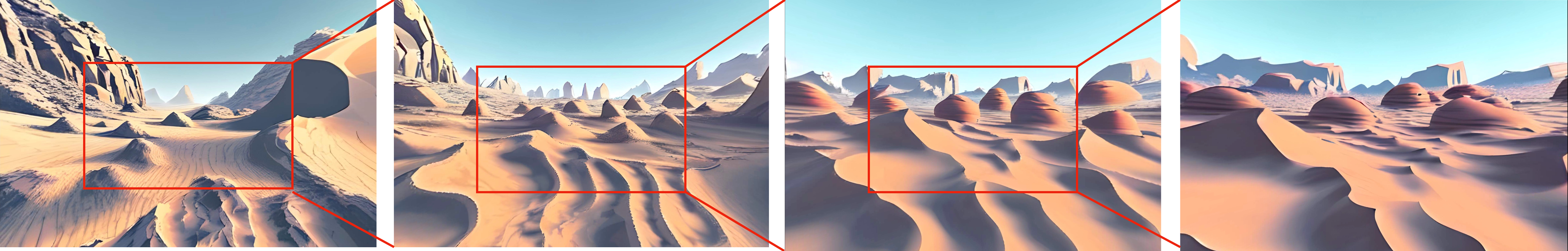}
  \caption{Example of a grid of scenes with forward movement between them. When combined, such scenes create an immersive experience of movement in the virtual world.}
\end{figure}

\section{Ethical Implications}
Ethical considerations in the realm of image generation hold paramount importance, particularly given the substantial implications of this technology. The ethical dimensions encompass a broad spectrum, particularly focusing on the potential utilization of such technologies to accentuate societal inequalities and proliferate visual representations entrenched in stereotypes and biases. The ramifications of deploying algorithms without careful consideration of their inherent biases and potentially harmful impacts can be profound, perpetuating existing disparities and possibly creating new ones.

\bibliography{references}

\newpage

\section{Appendix}
Let's define our parameters as: 
\begin{description}
  \item[step] The step of displacement, where \(0\) is no displacement, and \(1\) is displacement by the full radius.
  \item[direction] The direction of displacement in the form of an X-axis angle when looking at the sphere from above, an angle from \(0\) to \(359\) degrees.
  \item[width / height] The dimensions of the image.
\end{description}

To get the \((x\) coordinate of the point ``$b$'', we'll use next formulas:

\[ x_b = \left( \frac{{\text{{direction}}}}{360} \cdot \text{{width}} + \alpha \cdot \frac{{\text{{width}}}}{2} \right) \mod \text{{width}} \]
\[\alpha = \frac{{\text{{diff\_x\_norm}} - \arcsin \left( 0.5 \cdot \sin \left( \text{{diff\_x\_norm}} \right) \right)}}{\pi}\]
\[\text{{diff\_x\_norm}} = \frac{{2\pi \left( x - \frac{{\text{{direction}}}}{360} \cdot \text{{width}} \right)}}{\text{{width}}}\]

To get the \(y\) coordinate of the point ``$b$'', we'll use next formulas:
\[y_b = \frac{{\text{{width}} \cdot \left( \frac{1}{2} + \frac{{\beta}}{\pi} \right)}}{\pi}\]

We have two ``$\beta$'': one for the case when point ``$b$'' has crossed the zenith during the movement, and the second if it has crossed the nadir:
\[\beta_z = \text{{sign}} \left( \pi \frac{{y}}{\text{{height}}} - \frac{1}{2} \right) \cdot \left( \pi - \arccos \left( \frac{{\text{{step\_adjstep}} \cdot \alpha - \cos \text{{\_va}}}}{\sqrt{\text{{step}} \cdot \alpha^2 - 2 \cdot \text{{step}} \cdot \alpha \cdot \cos \text{{\_va}} + 1}} \right) \right)\]

\[\beta_n = \text{{sign}} \left( \pi \frac{{y}}{\text{{height}}} - \frac{1}{2} \right) \cdot \arccos \left( -\frac{{\text{{step}} \cdot \alpha - \cos \left( \pi \frac{{y}}{\text{{height}}} - \frac{1}{2} \right)}}{\sqrt{\text{{step}} \cdot \alpha^2 - 2 \cdot \text{{step}} \cdot \alpha \cdot \cos \left( \pi  \frac{{y}}{\text{{height}}} - \frac{1}{2} \right) + 1}} \right)\]

\[ \text{{step\_adjusted}} = \text{{step}} \cdot \alpha \]


\begin{figure}[htbp]
    \centering
    \begin{subfigure}{0.8\textwidth}
        \centering
        \includegraphics[width=\linewidth]{initial_scene.png}
        \caption{Initial panorama}
        \vspace{1cm}
    \end{subfigure}
    \begin{subfigure}{0.8\textwidth}
        \centering
        \includegraphics[width=\linewidth]{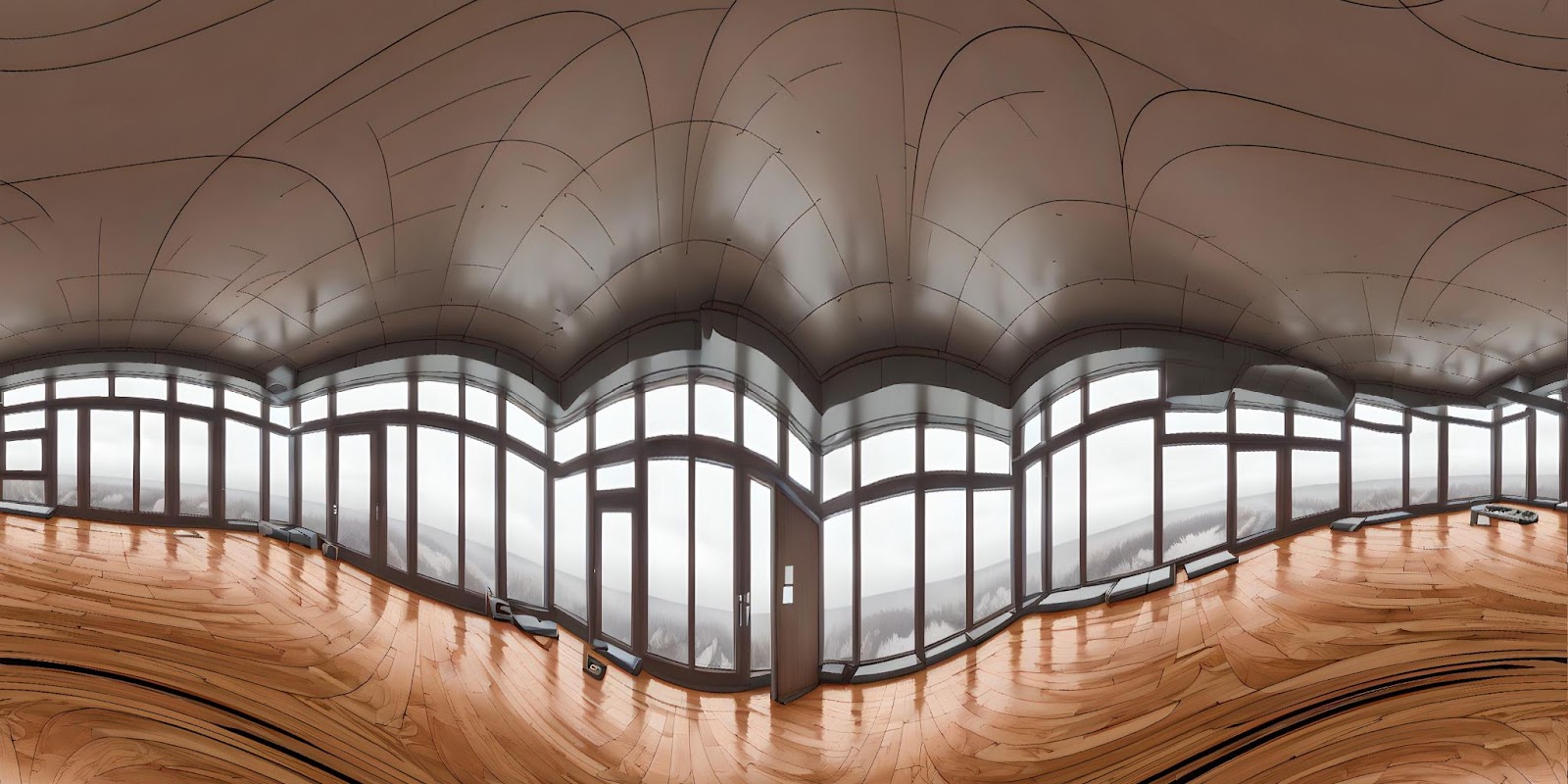}
        \caption{Distorted panorama}
        \vspace{1cm}
    \end{subfigure}
    \hfill
    \begin{subfigure}{0.8\textwidth}
        \centering
        \includegraphics[width=\linewidth]{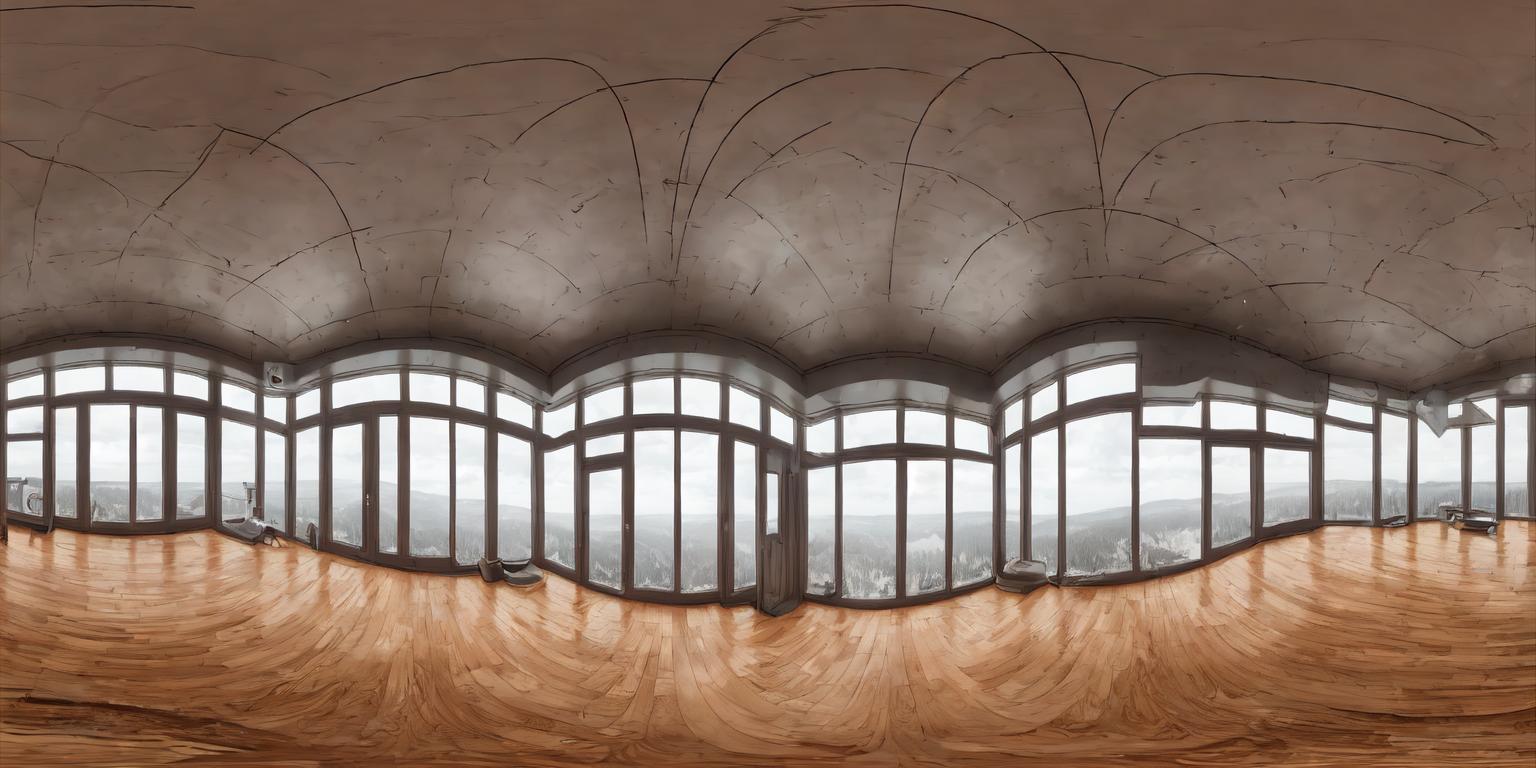}
        \caption{Restored panorama after distortion}
        \vspace{1cm}
    \end{subfigure}
    \caption{Process of the panorama restoration.}
    \label{fig:three_images}
\end{figure}

\begin{figure}[htbp]
    \centering
    \begin{subfigure}{0.8\textwidth}
        \centering
        \includegraphics[width=\linewidth]{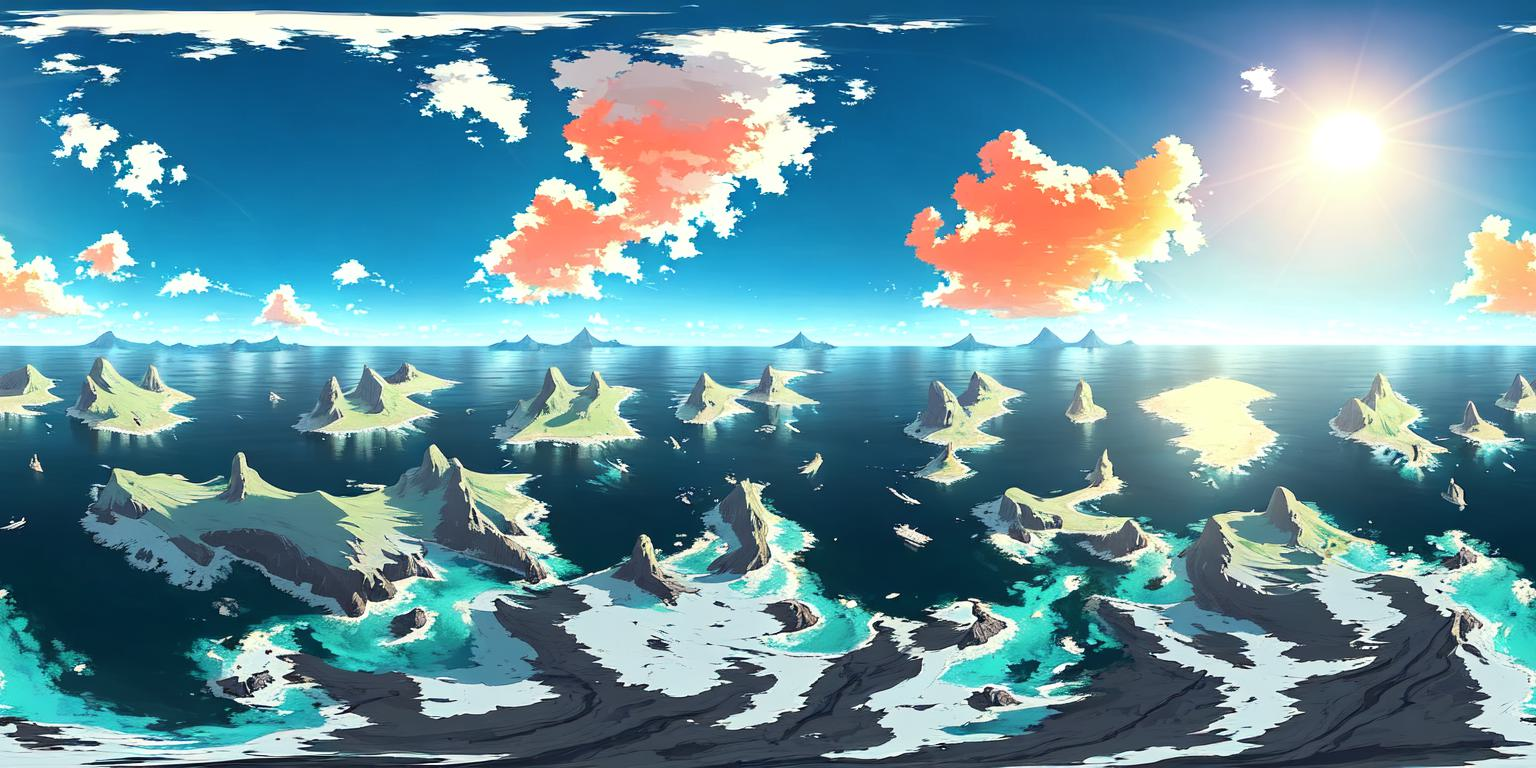}
        \caption{Initial panorama}
        \vspace{1cm}
    \end{subfigure}
    \hfill
    \begin{subfigure}{0.8\textwidth}
        \centering
        \includegraphics[width=\linewidth]{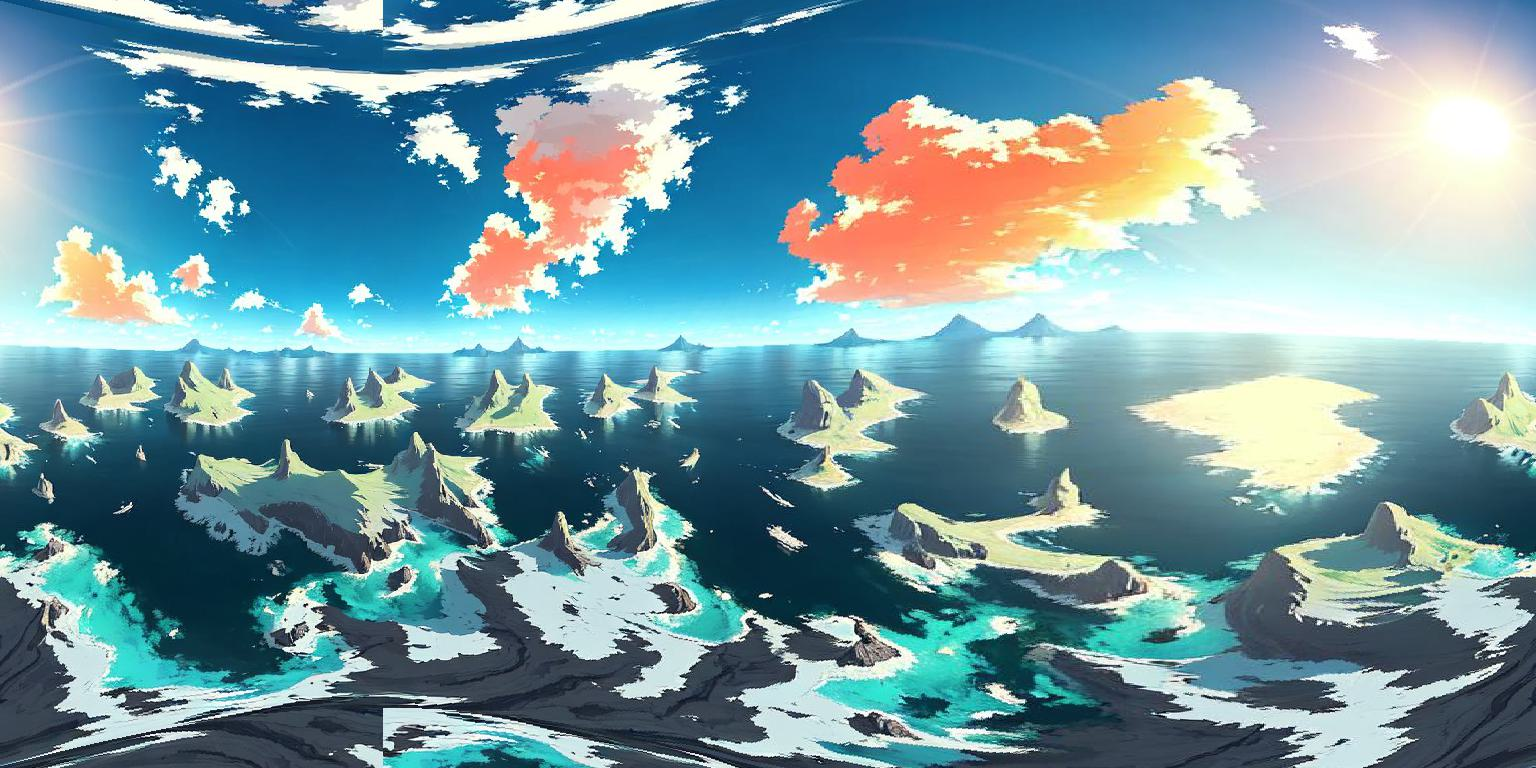}
        \caption{Distorted panorama}
        \vspace{1cm}
    \end{subfigure}
    \hfill
    \begin{subfigure}{0.8\textwidth}
        \centering
        \includegraphics[width=\linewidth]{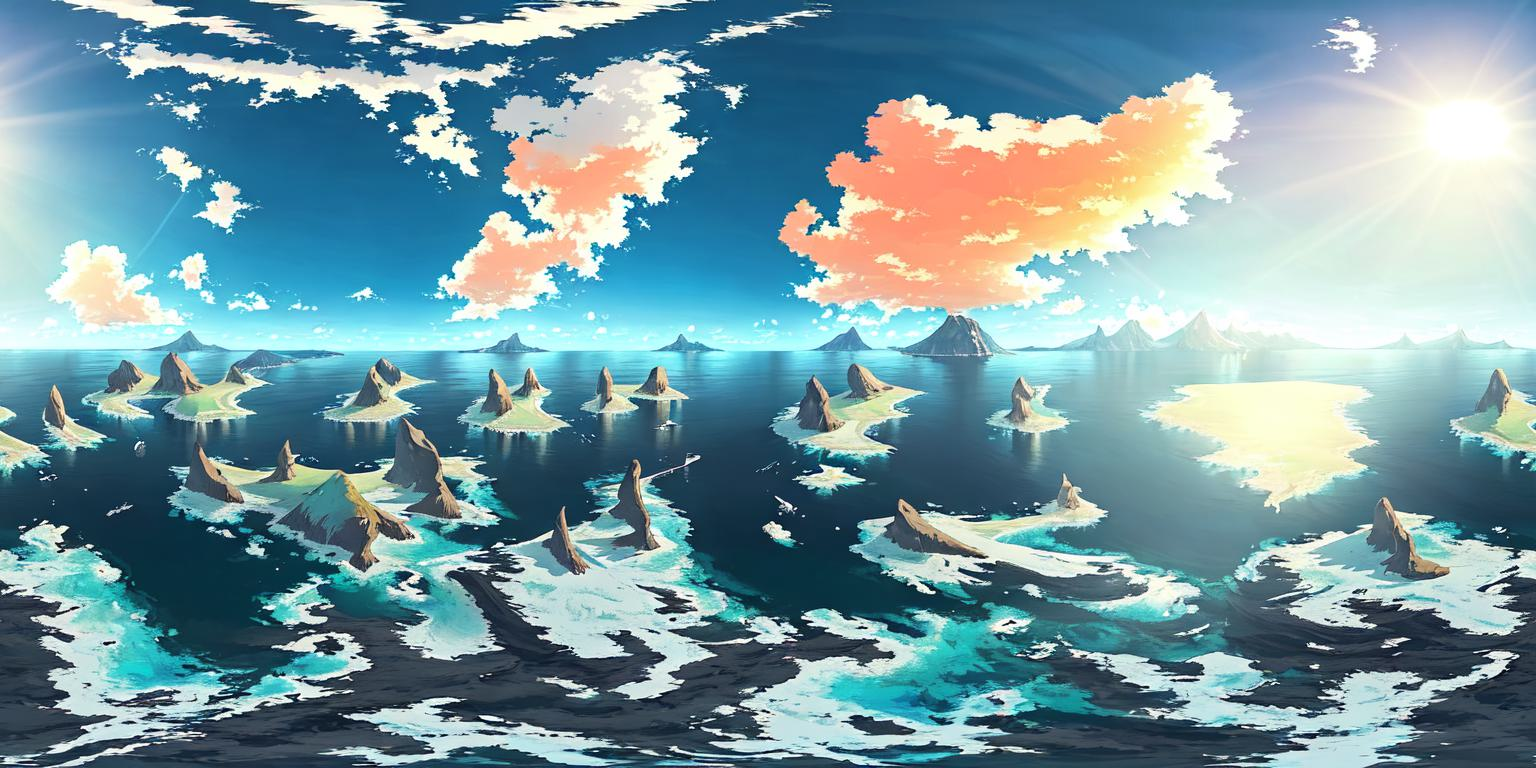}
        \caption{Restored panorama after distortion}
        \vspace{1cm}
    \end{subfigure}
    \caption{Process of the panorama restoration.}
    \label{fig:three_images}
\end{figure}

\begin{figure}[htbp]
    \centering
    \begin{subfigure}{0.45\textwidth}
        \centering
        \includegraphics[width=\linewidth]{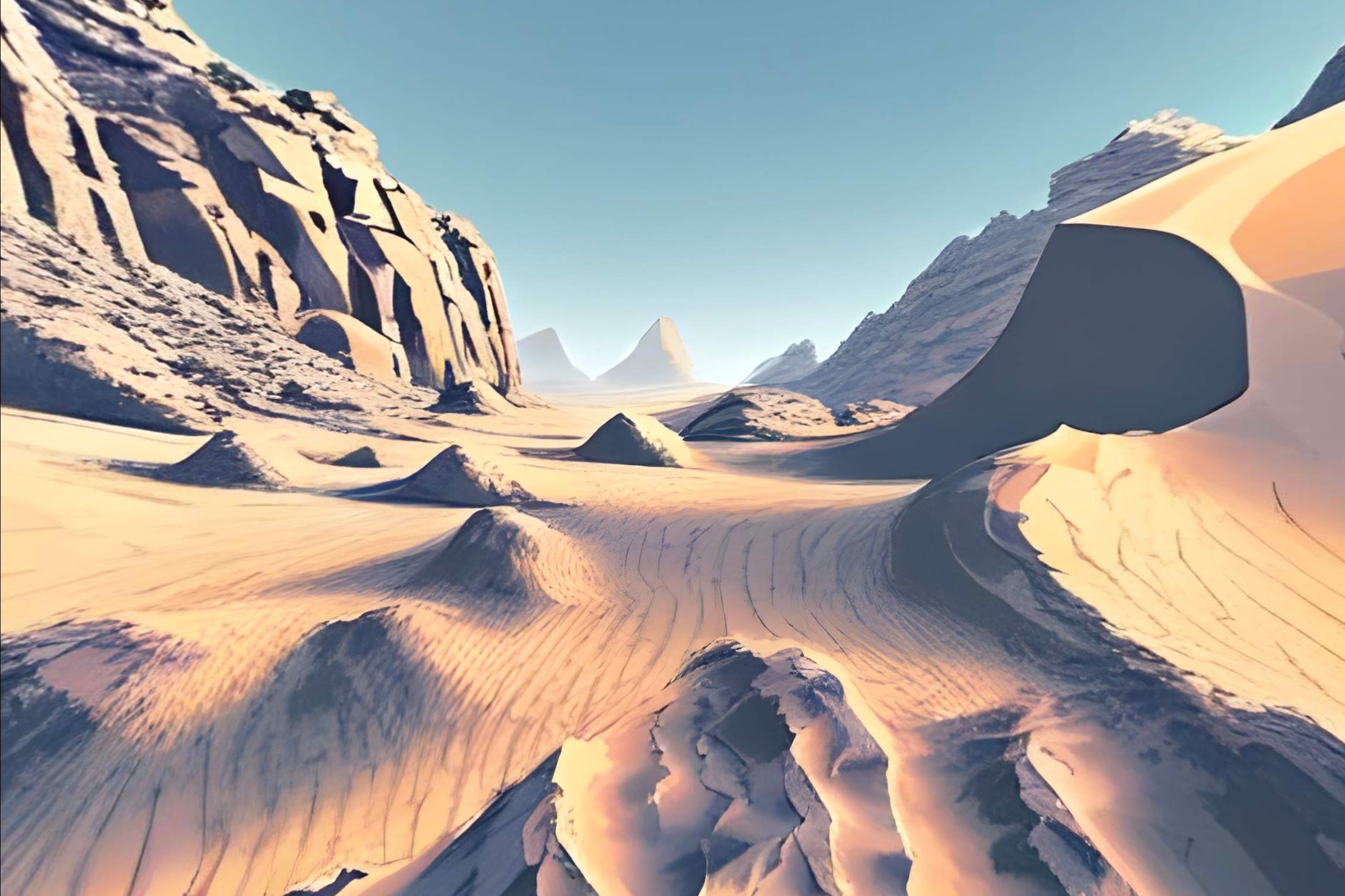}
        \caption{Scene 1}
        \vspace{0.5cm}
    \end{subfigure}
    \hfill
    \begin{subfigure}{0.45\textwidth}
        \centering
        \includegraphics[width=\linewidth]{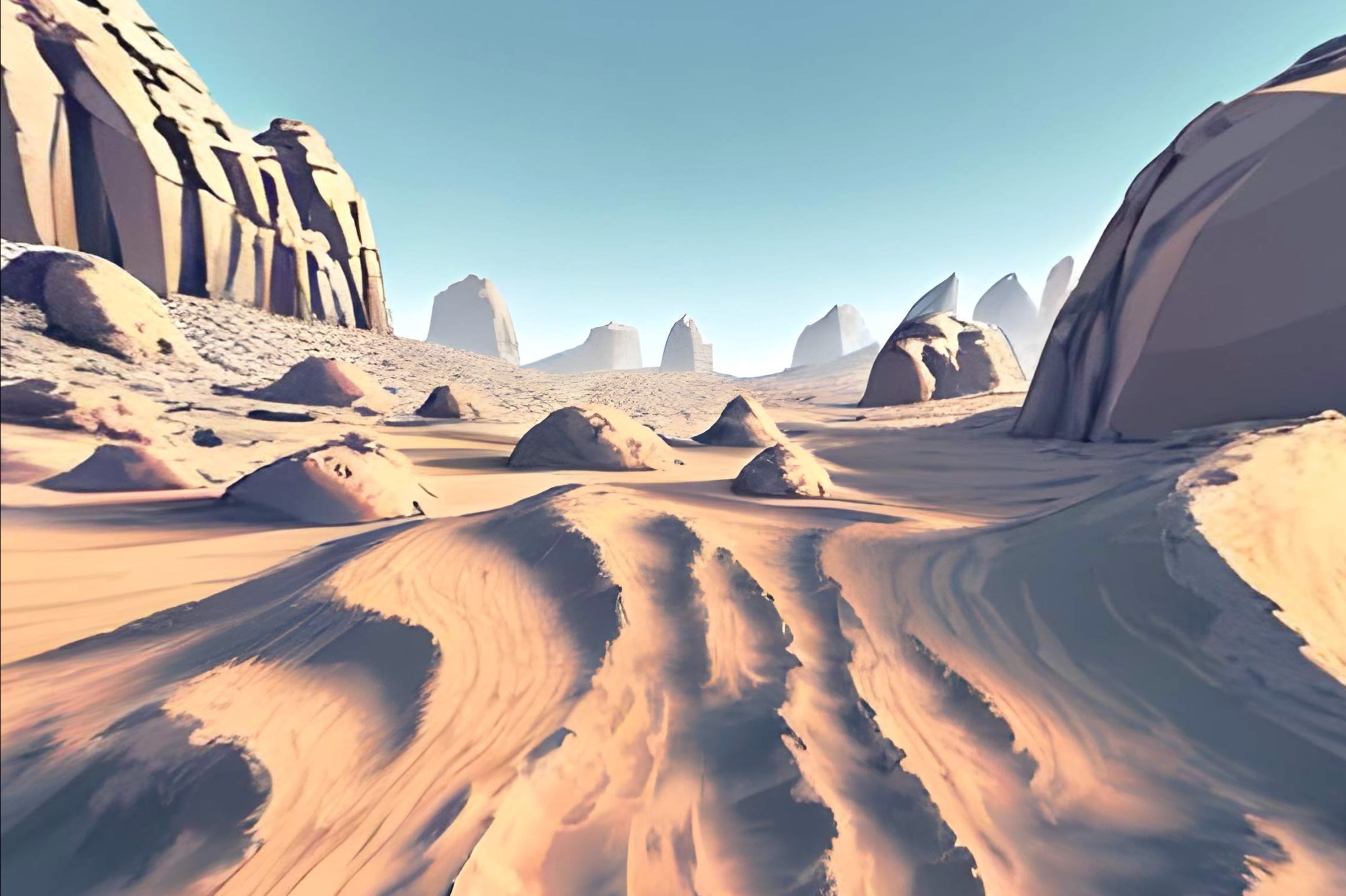}
        \caption{Scene 2}
        \vspace{0.5cm}
    \end{subfigure}
    \hfill
    \begin{subfigure}{0.45\textwidth}
        \centering
        \includegraphics[width=\linewidth]{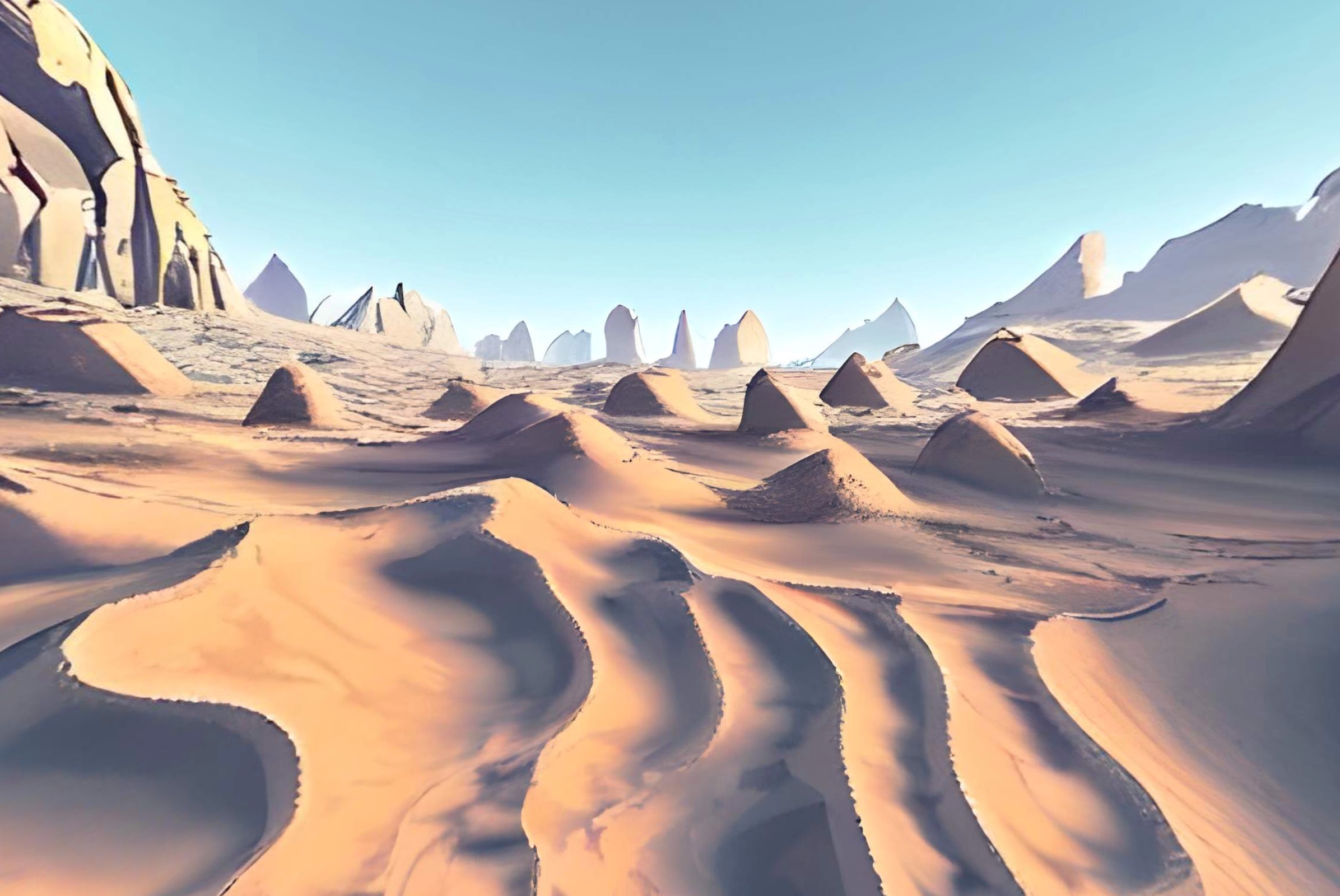}
        \caption{Scene 3}
        \vspace{0.5cm}
    \end{subfigure}
        \hfill
    \begin{subfigure}{0.45\textwidth}
        \centering
        \includegraphics[width=\linewidth]{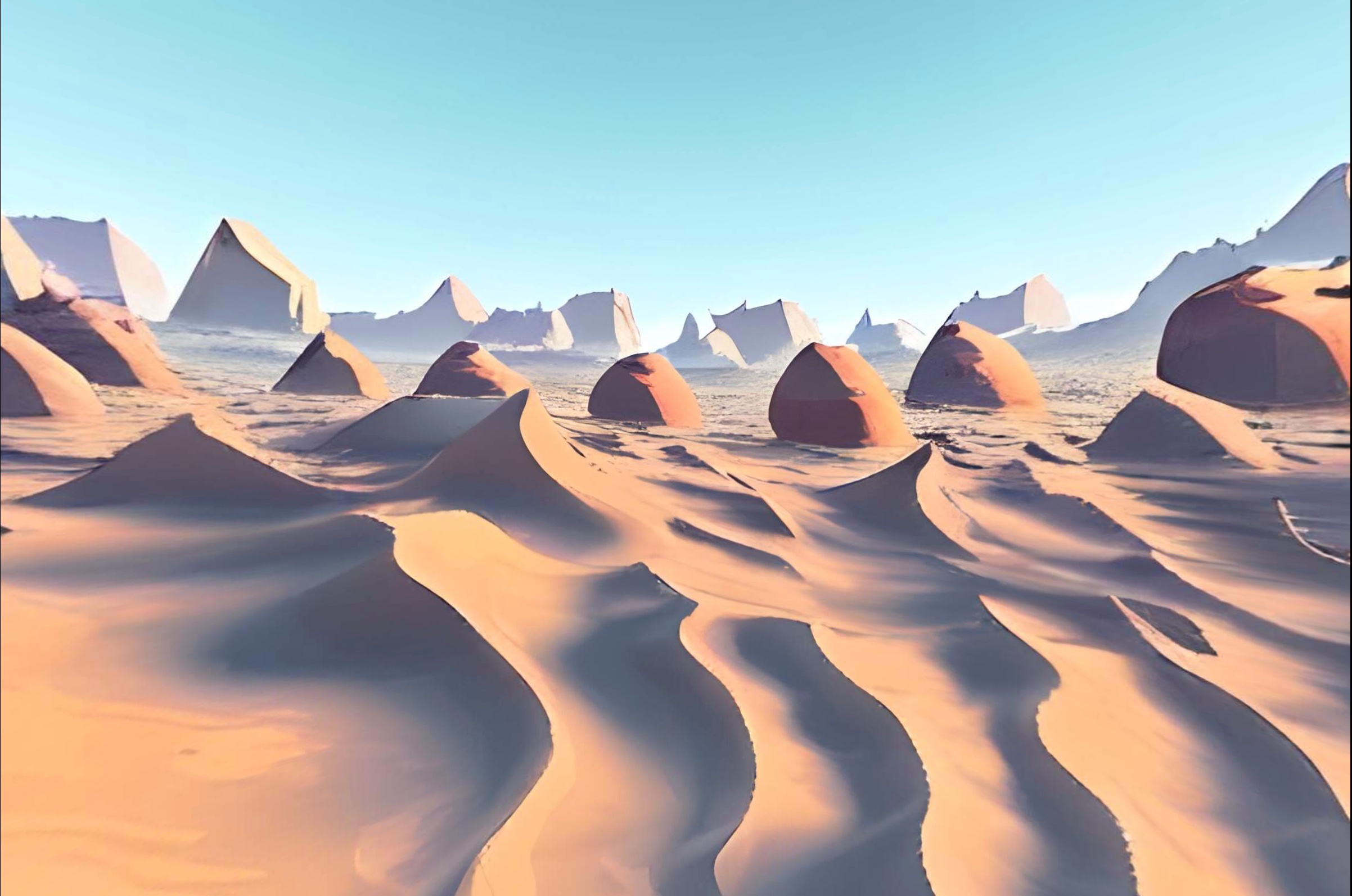}
        \caption{Scene 4}
        \vspace{0.5cm}
    \end{subfigure}
        \hfill
    \begin{subfigure}{0.45\textwidth}
        \centering
        \includegraphics[width=\linewidth]{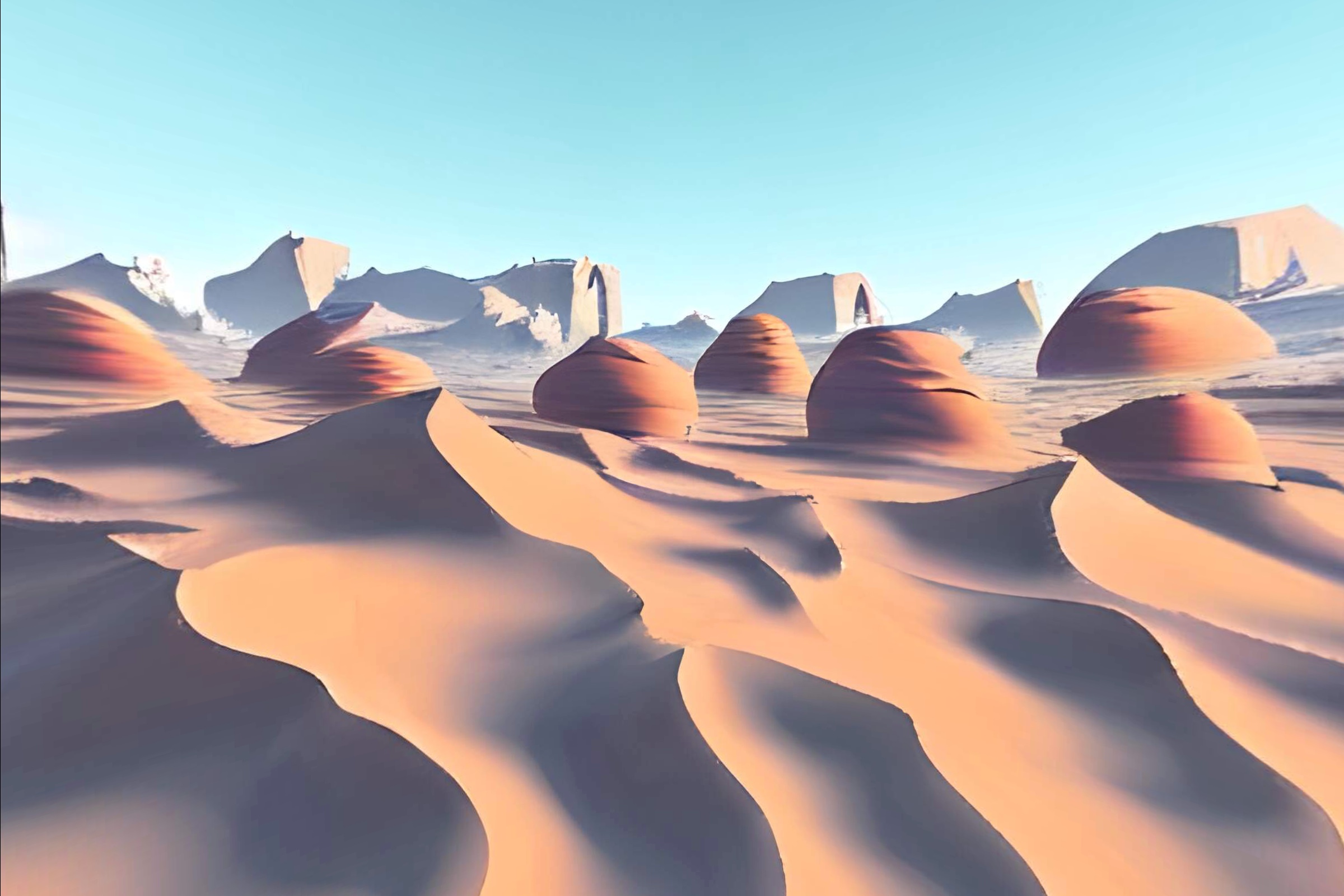}
        \caption{Scene 5}
        \vspace{0.5cm}
    \end{subfigure}
        \hfill
    \begin{subfigure}{0.45\textwidth}
        \centering
        \includegraphics[width=\linewidth]{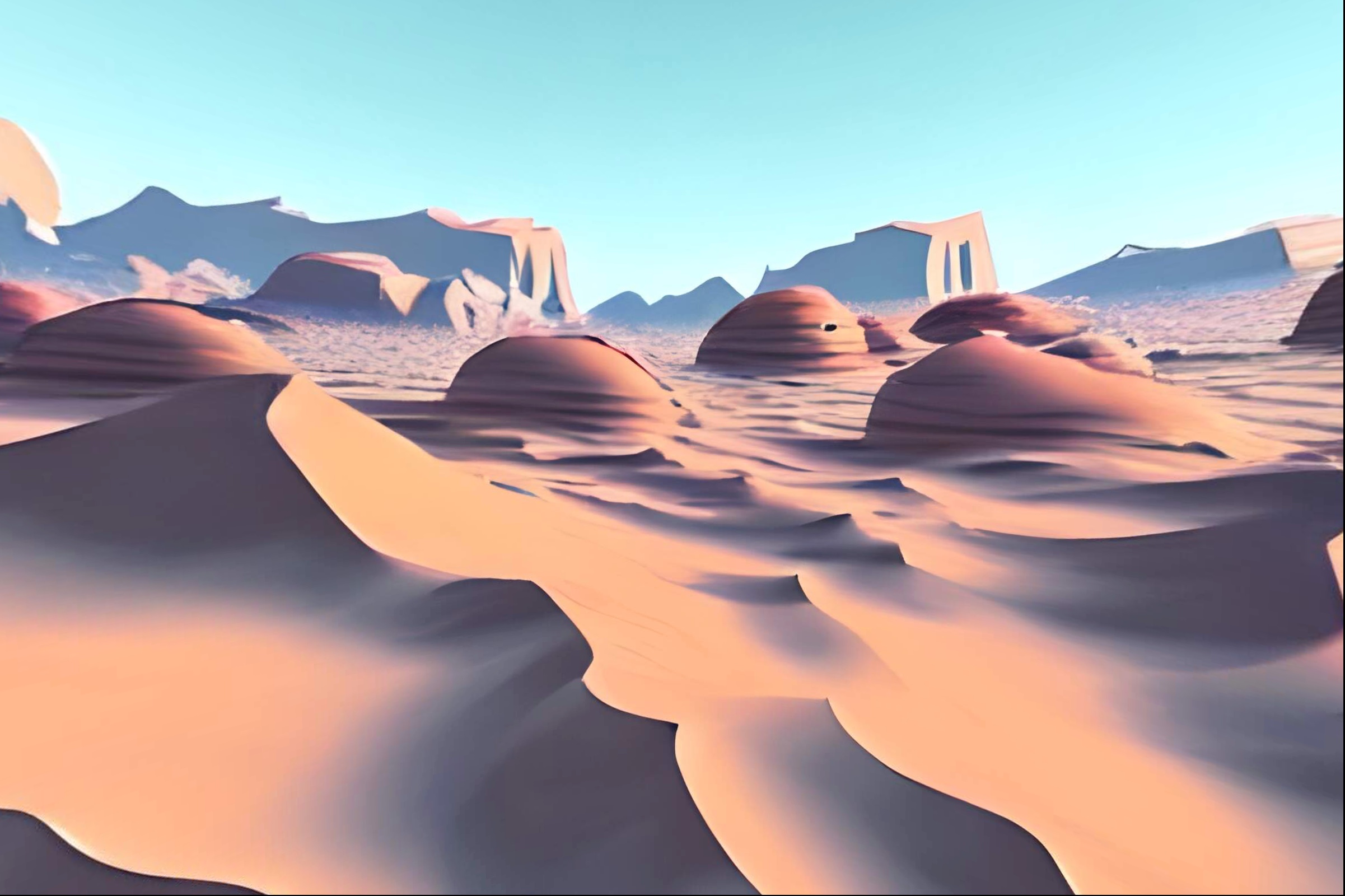}
        \caption{Scene 6}
        \vspace{0.5cm}
    \end{subfigure}
    \caption{A simple desert world, which consists of 6 scenes with the structure 1->2->3->4->5->6.}
    \label{fig:three_images}
\end{figure}

\end{document}


\maketitle

\section{Appendix}

To get the \((x, y)\) coordinates of the point ``$b$'', we'll use next formula: 

\begin{align}
x_b = \left(\frac{{\text{{direction}}}}{360} \cdot \text{{width}} + \alpha \cdot \frac{{\text{{width}}}}{2} \right) \mod \text{{width}} \\
\text{{where}} \\
 \alpha = \frac{{2\pi \left( x - \frac{{\text{{direction}}}}{360} \cdot \text{{width}} \right) - \arcsin \left( 0.5 \cdot \sin \left( 2\pi \left( x - \frac{{\text{{direction}}}}{360} \cdot \text{{width}} \right) / \text{{width}} \right) \right)}}{\pi} \\
\end{align}

\begin{align}
y_b = \frac{{\text{{width}} \left( \frac{1}{2} + \beta \right)}}{\pi} \\
\text{{where}} \\
\beta = \text{{sign}} \left( \pi \cdot \left( \frac{{y}}{\text{{height}}} - \frac{1}{2} \right) \right) \cdot \left( \pi - \arccos \left( \frac{{\text{{step}} \cdot \alpha - \cos \left( \pi \cdot \left( \frac{{y}}{\text{{height}}} - \frac{1}{2} \right) \right)}}{\sqrt{\text{{step}} \cdot \alpha^2 - 2 \cdot \text{{step}} \cdot \alpha \cdot \cos \left( \pi \cdot \left( \frac{{y}}{\text{{height}}} - \frac{1}{2} \right) \right) + 1}} \right) \right) \\
\beta = \text{{sign}} \left( \pi \cdot \left( \frac{{y}}{\text{{height}}} - \frac{1}{2} \right) \right) \cdot \arccos \left( -\frac{{\text{{step}} \cdot \alpha - \cos \left( \pi \cdot \left( \frac{{y}}{\text{{height}}} - \frac{1}{2} \right) \right)}}{\sqrt{\text{{step}} \cdot \alpha^2 - 2 \cdot \text{{step}} \cdot \alpha \cdot \cos \left( \pi \cdot \left( \frac{{y}}{\text{{height}}} - \frac{1}{2} \right) \right) + 1}} \right)
\end{align}


\begin{figure}[htbp]
    \centering
    \begin{subfigure}{0.32\textwidth}
        \centering
        \includegraphics[width=\linewidth]{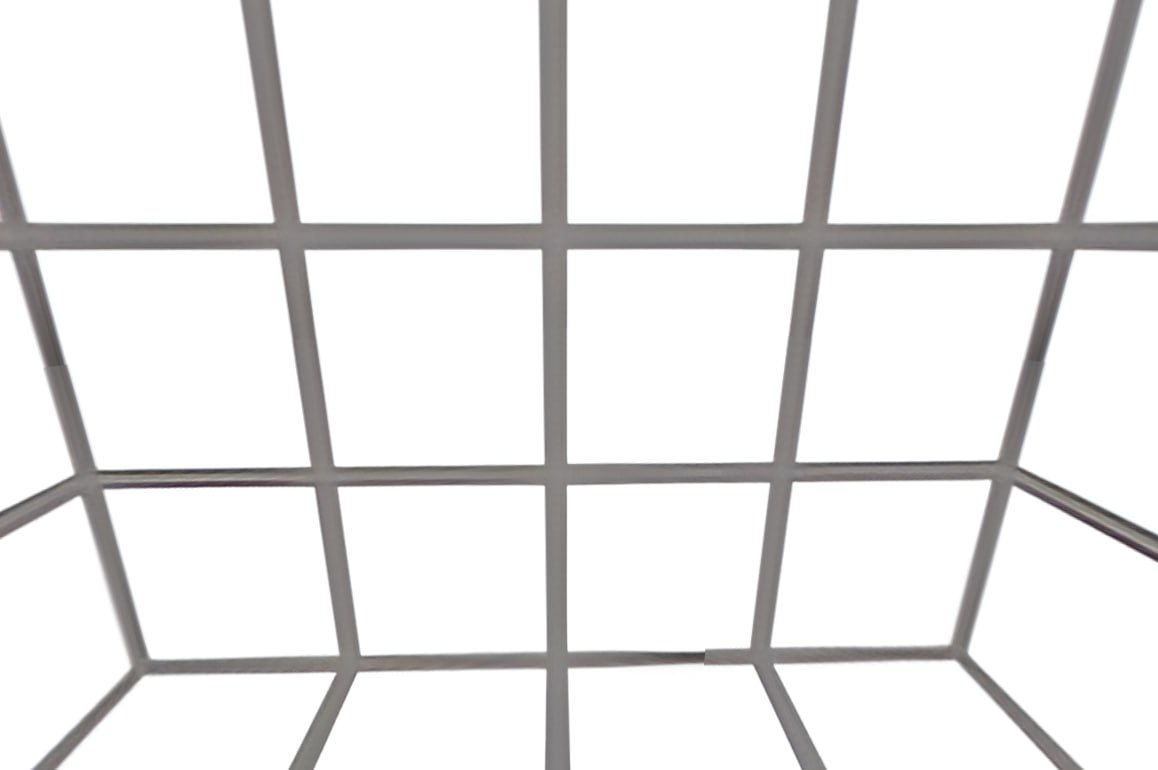}
        \caption{Initial panorama}
    \end{subfigure}
    \hfill
    \begin{subfigure}{0.32\textwidth}
        \centering
        \includegraphics[width=\linewidth]{distorted_projection.jpg}
        \caption{Distorted panorama}
    \end{subfigure}
    \hfill
    \begin{subfigure}{0.32\textwidth}
        \centering
        \includegraphics[width=\linewidth]{restored_projection.jpg}
        \caption{Restored after distortion}
    \end{subfigure}
    \caption{Process of the panorama restoration}
    \label{fig:three_images}
\end{figure}

\begin{figure}[htbp]
    \centering
    \begin{subfigure}{0.8\textwidth}
        \centering
        \includegraphics[width=\linewidth]{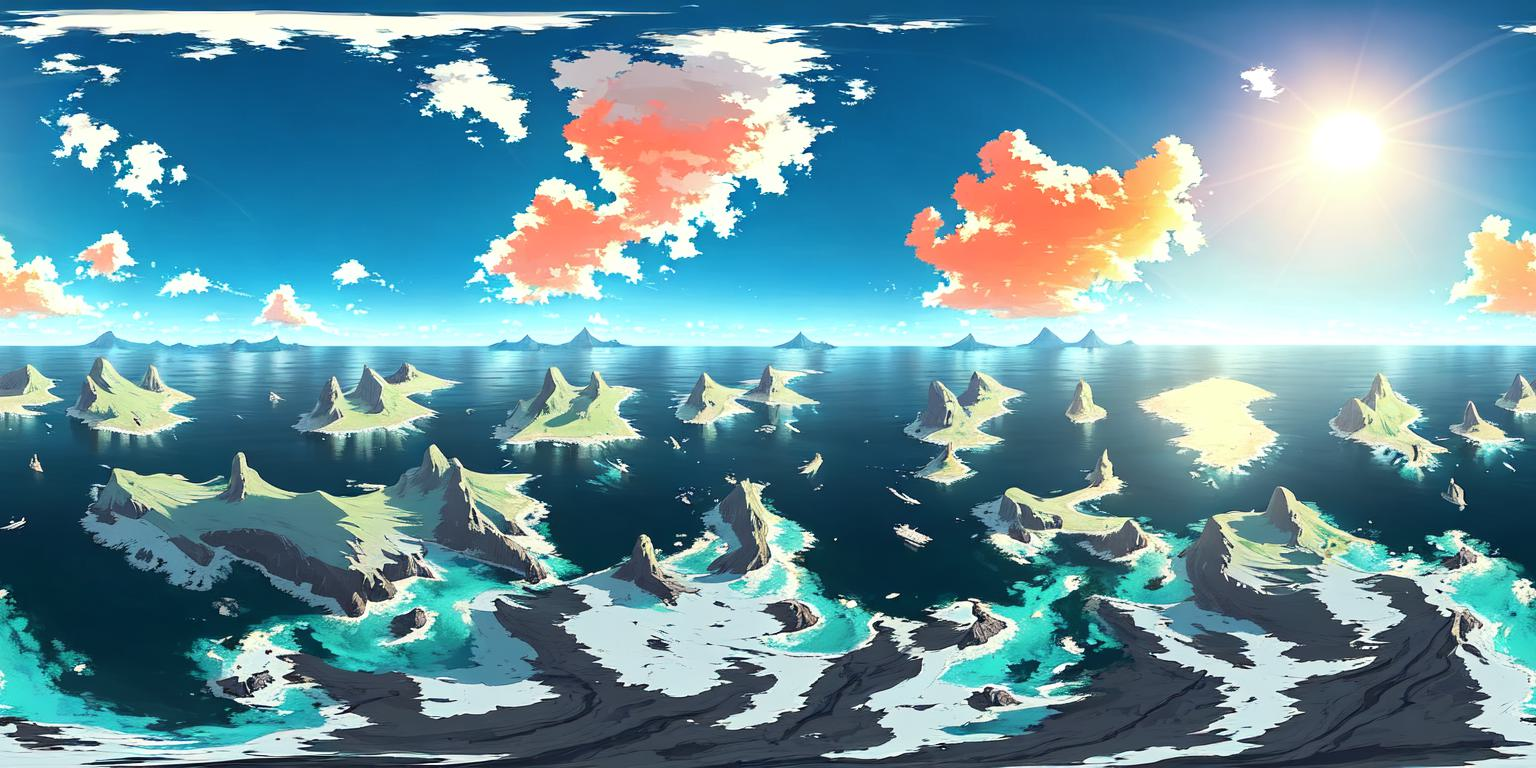}
        \caption{Initial panorama}
    \end{subfigure}
    \hfill
    \begin{subfigure}{0.8\textwidth}
        \centering
        \includegraphics[width=\linewidth]{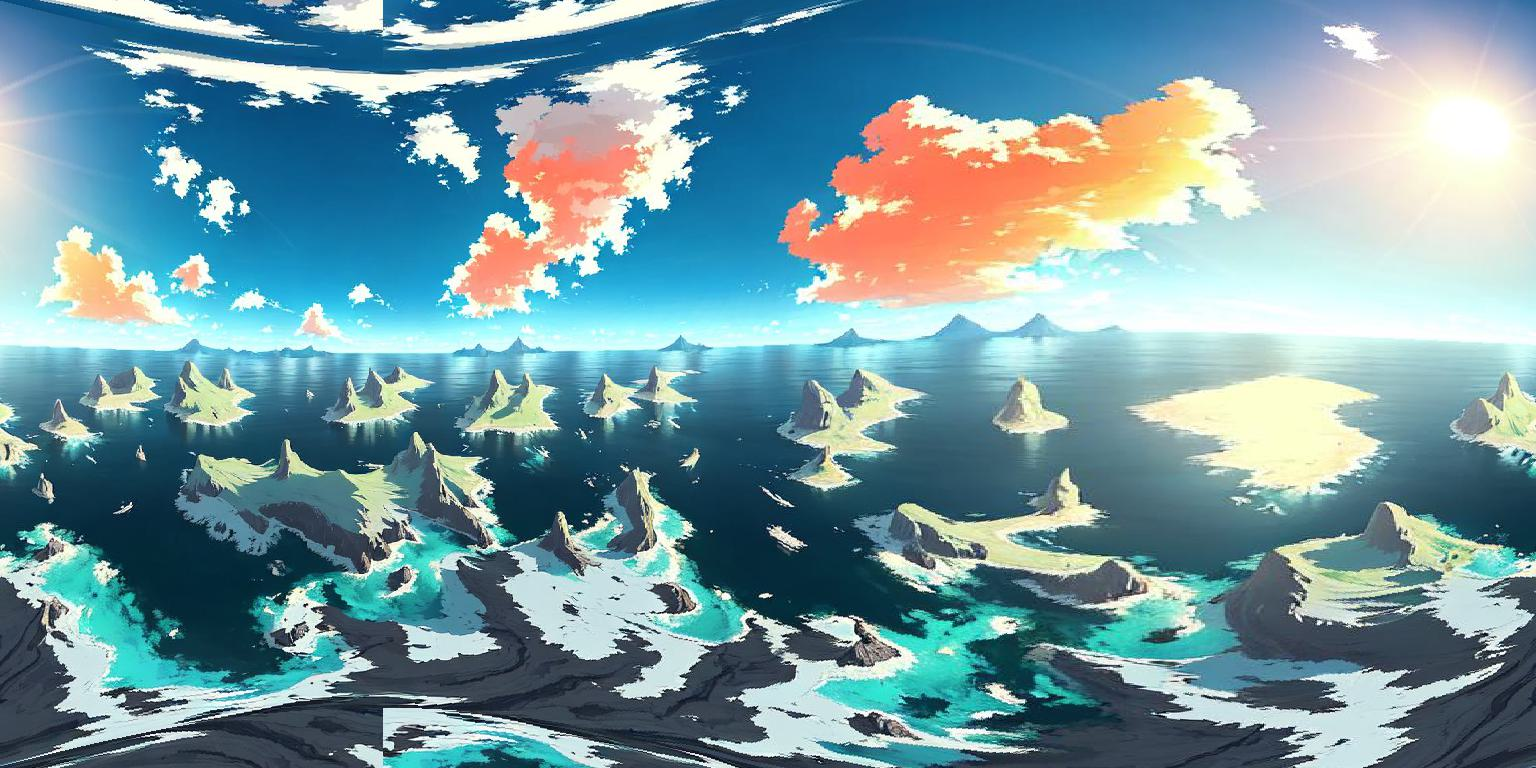}
        \caption{Distorted panorama}
    \end{subfigure}
    \hfill
    \begin{subfigure}{0.8\textwidth}
        \centering
        \includegraphics[width=\linewidth]{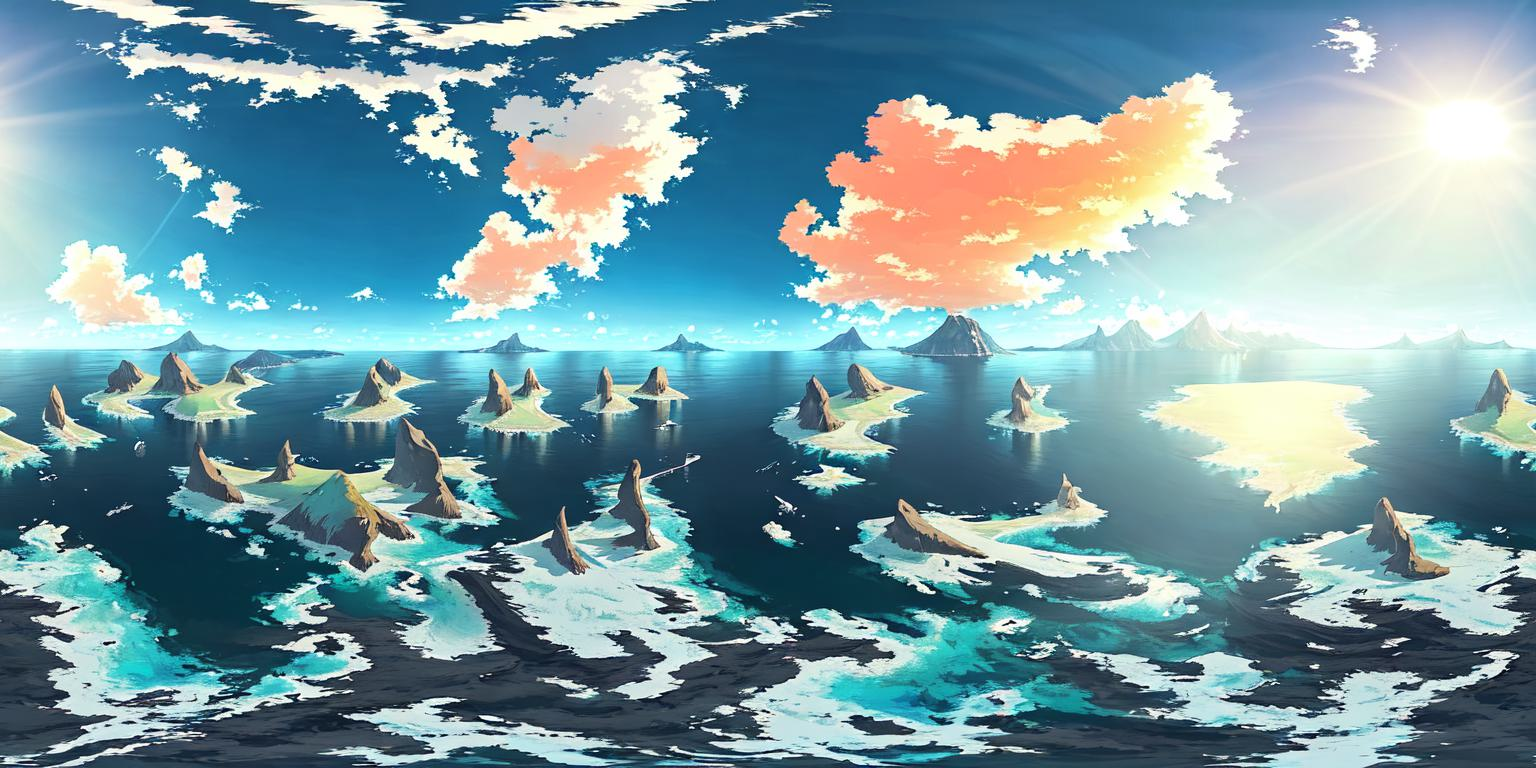}
        \caption{Restored after distortion}
    \end{subfigure}
    \caption{Process of the panorama restoration}
    \label{fig:three_images}
\end{figure}

\begin{figure}[htbp]
    \centering
    \begin{subfigure}{0.45\textwidth}
        \centering
        \includegraphics[width=\linewidth]{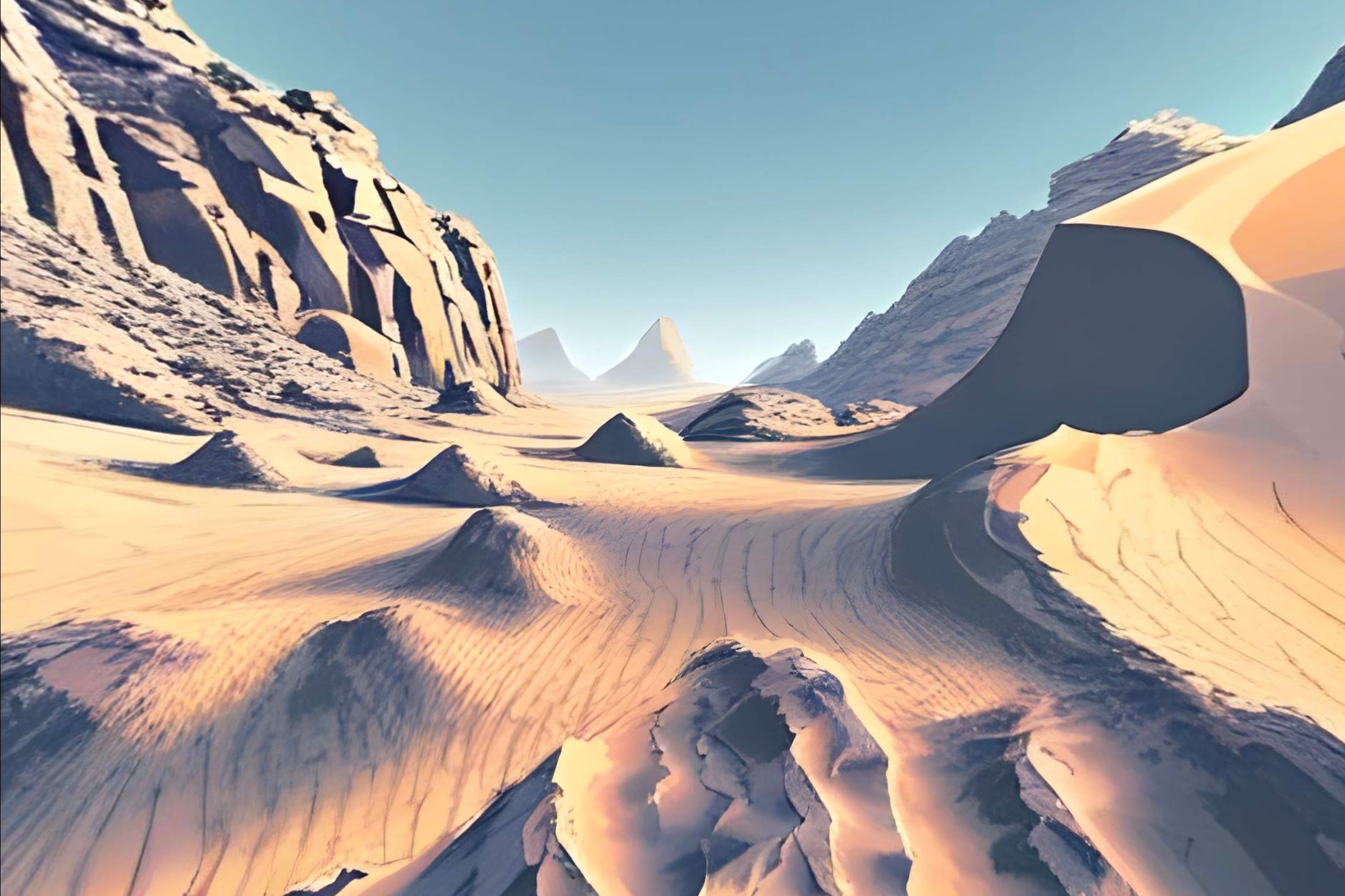}
        \caption{1}
    \end{subfigure}
    \hfill
    \begin{subfigure}{0.45\textwidth}
        \centering
        \includegraphics[width=\linewidth]{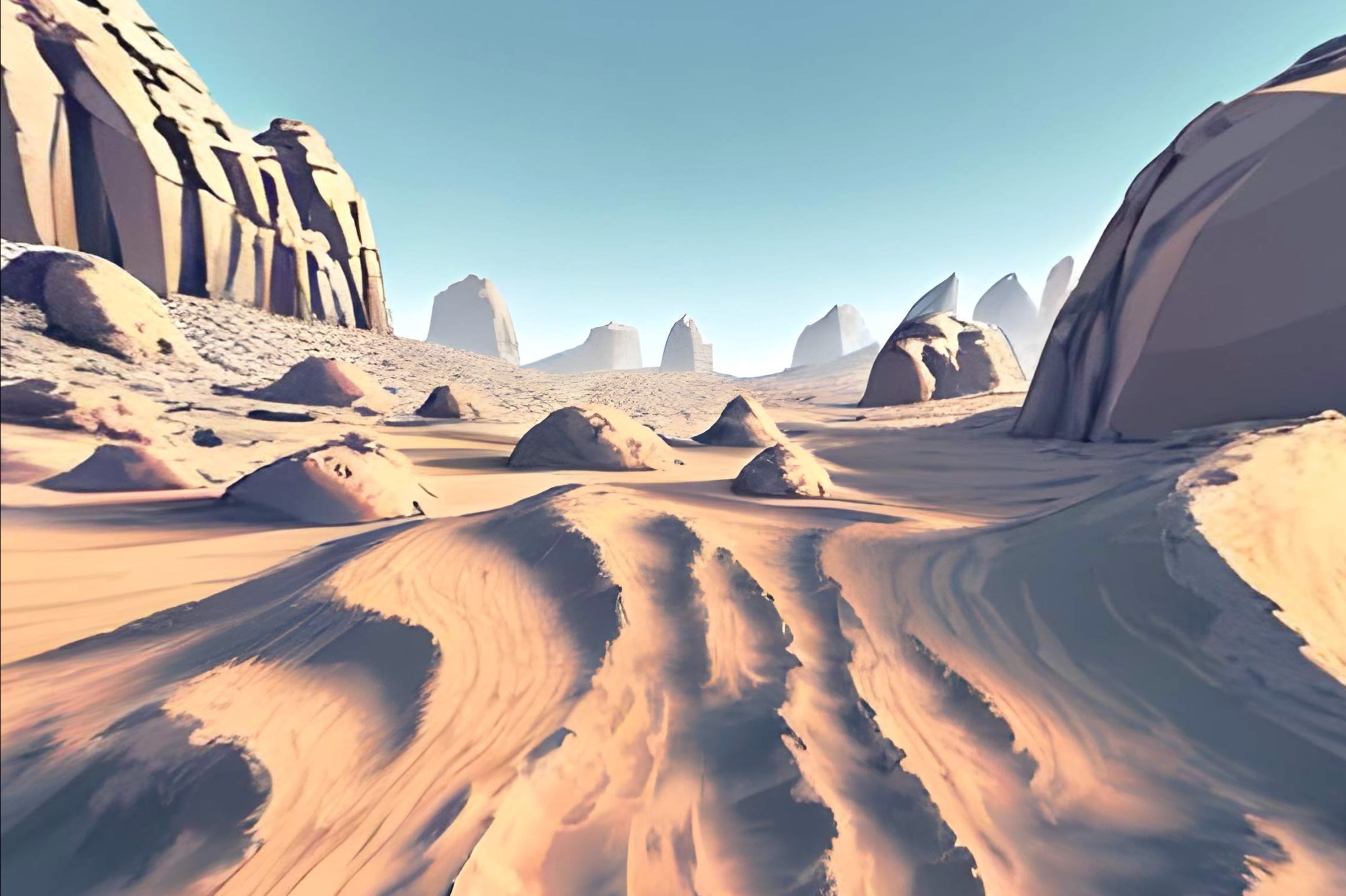}
        \caption{2}
    \end{subfigure}
    \hfill
    \begin{subfigure}{0.45\textwidth}
        \centering
        \includegraphics[width=\linewidth]{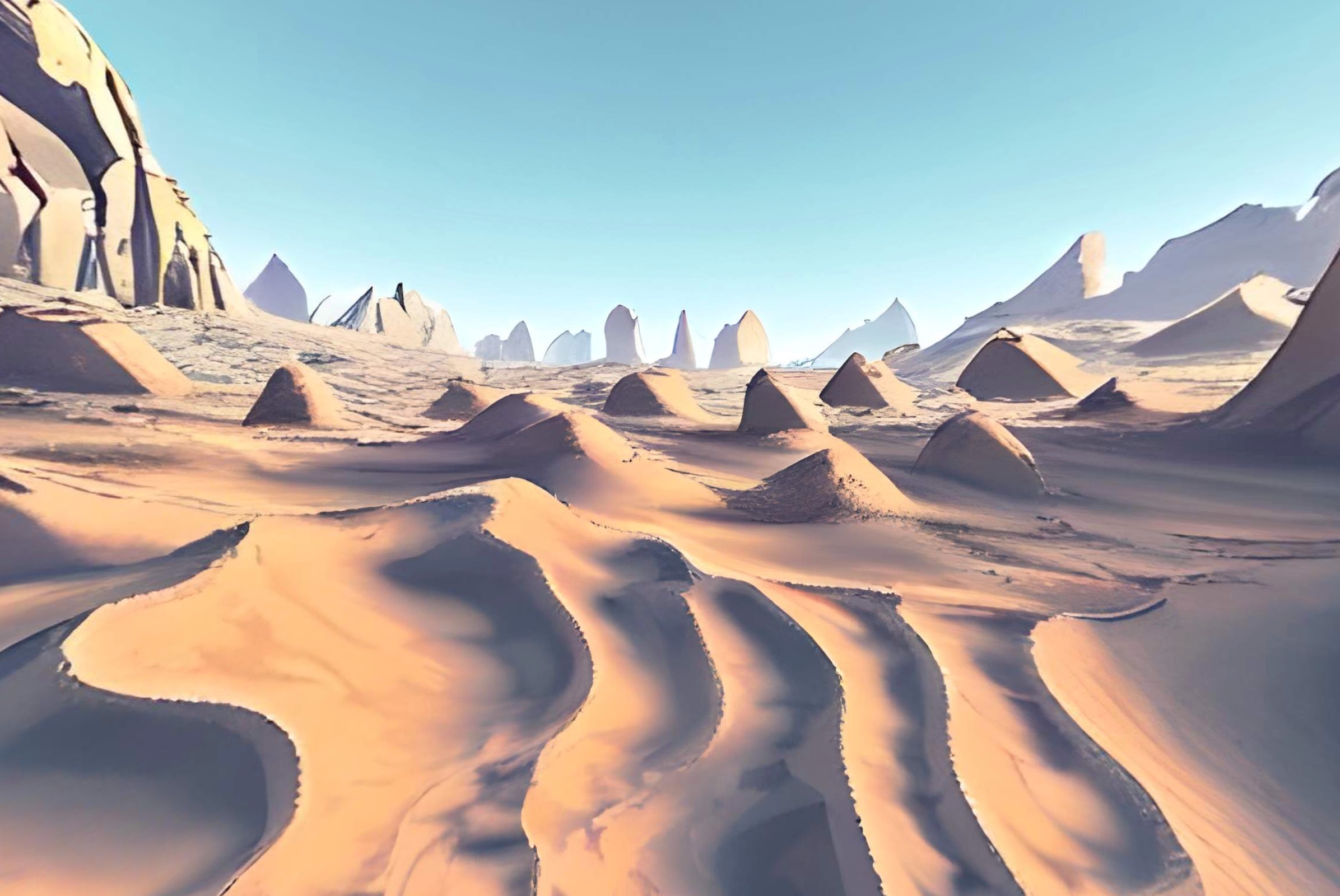}
        \caption{3}
    \end{subfigure}
        \hfill
    \begin{subfigure}{0.45\textwidth}
        \centering
        \includegraphics[width=\linewidth]{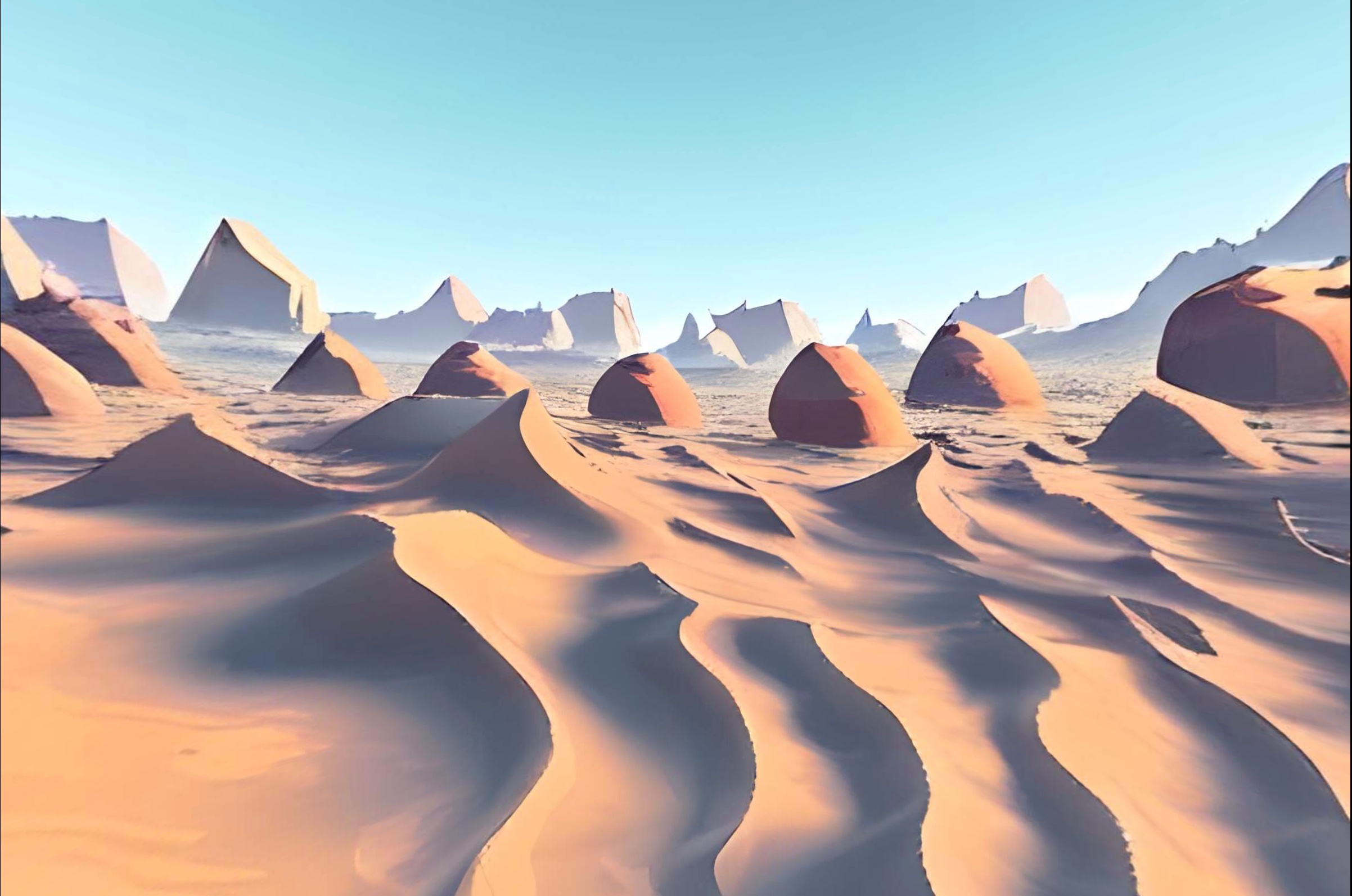}
        \caption{4}
    \end{subfigure}
        \hfill
    \begin{subfigure}{0.45\textwidth}
        \centering
        \includegraphics[width=\linewidth]{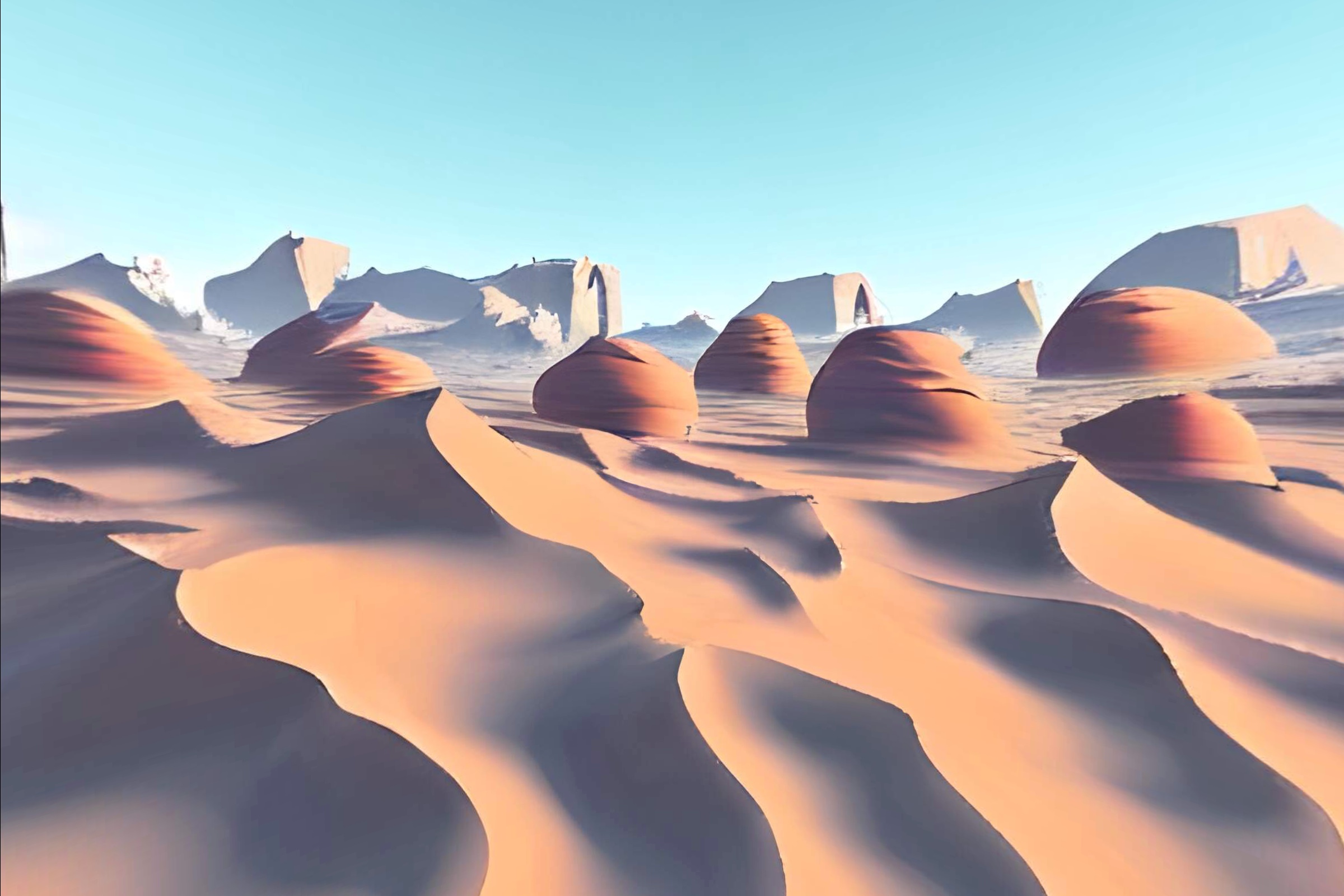}
        \caption{5}
    \end{subfigure}
        \hfill
    \begin{subfigure}{0.45\textwidth}
        \centering
        \includegraphics[width=\linewidth]{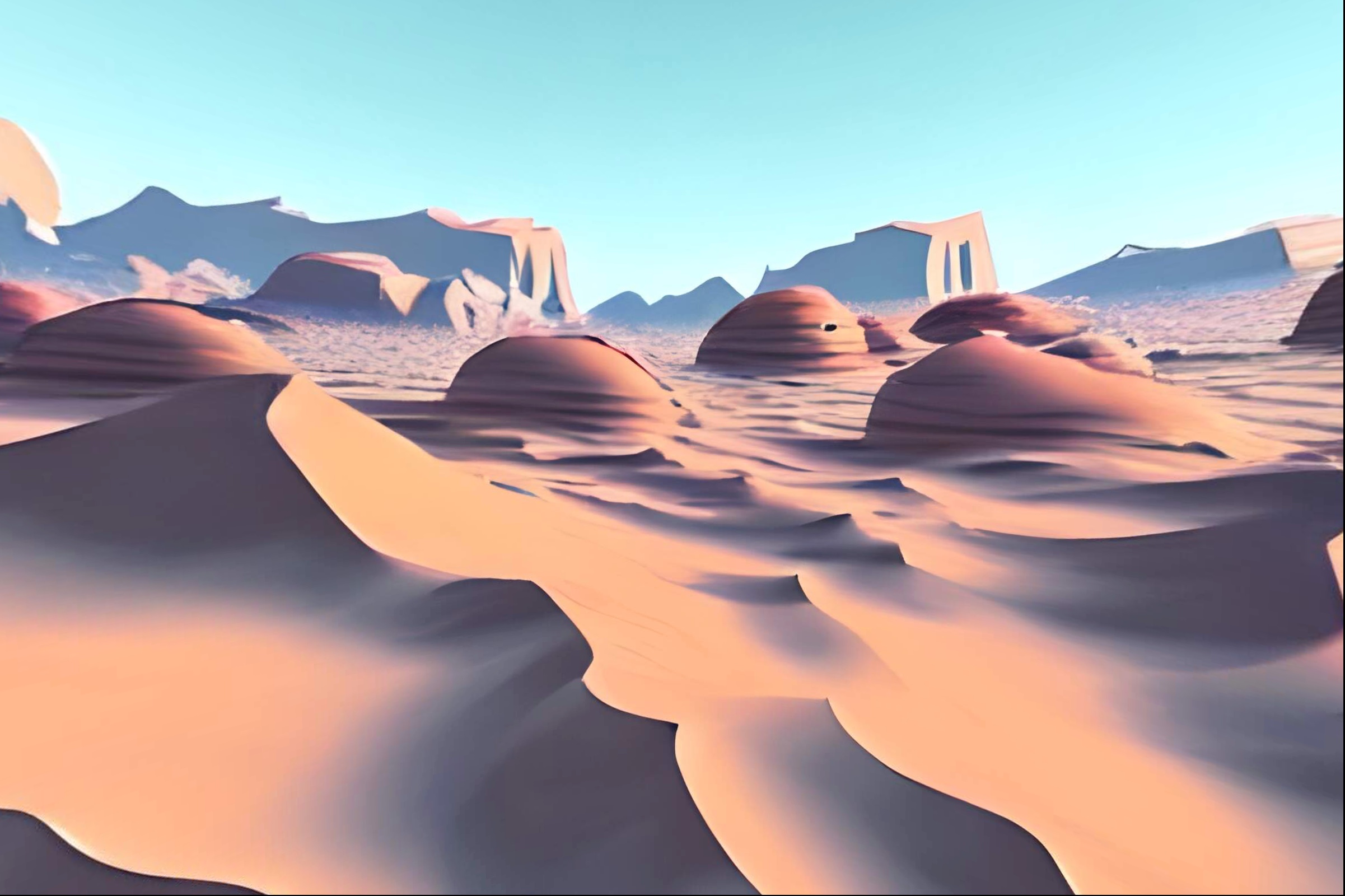}
        \caption{6}
    \end{subfigure}
    \caption{One of the worlds}
    \label{fig:three_images}
\end{figure}